\documentclass[prd,preprint,showpacs,preprintnumbers,amsmath,amssymb,eqsecnum]{revtex4}

\usepackage{graphicx}
\usepackage{dcolumn}
\usepackage{bm}
\usepackage{subfigure}
\usepackage{color}

\begin{document}

\preprint{RUP 11-1}
\preprint{OCU-PHYS 346}
\preprint{AP-GR 89}

\title{
Collision of two general geodesic particles 
around a Kerr black hole
}
\author{$^{1}$Tomohiro Harada}%
 \email{harada@rikkyo.ac.jp}
\author{$^{2}$Masashi Kimura}
\email{mkimura@sci.osaka-cu.ac.jp}
\affiliation{%
$^{1}$Department of Physics, Rikkyo University, Toshima, Tokyo 175-8501, Japan\\
$^{2}$Department of Mathematics and Physics, Graduate School of Science,
Osaka City University, Osaka 558-8585, Japan
}%
\date{\today}

\begin{abstract}
We obtain an explicit expression for the center-of-mass (CM) energy  
of two colliding general geodesic 
massive and massless particles at any spacetime point 
around a Kerr black hole. 
Applying this, 
we show that the CM energy can be arbitrarily 
high only in the limit to the horizon and then 
derive a formula for the CM energy 
of two general geodesic particles colliding 
near the horizon
in terms of the conserved quantities of each particle and the 
polar angle.
We present the necessary and sufficient condition 
for the CM energy to be arbitrarily high
in terms of the conserved quantities of each particle.
To have an arbitrarily high CM energy, 
the angular momentum of either of the two particles 
must be fine-tuned to the critical value
$L_{i}=\Omega_{H}^{-1}E_{i}$, 
where $\Omega_{H}$ is the angular velocity of the horizon
and $E_{i}$ and $L_{i}$ 
are the energy and angular momentum of particle $i$ ($=1,2$), 
respectively.
We show that, in the direct collision scenario, 
the collision with an 
arbitrarily high CM energy can occur 
near the horizon of maximally rotating black holes not only at 
the equator but also on a belt centered at the equator. 
This belt lies between latitudes
$\pm \mbox{acos}(\sqrt{3}-1)\simeq \pm 42.94^{\circ}$.
This is also true in the scenario
through the collision of a last stable orbit particle.
\end{abstract}

\maketitle


\section{Introduction}
Ba\~nados, Silk and West~\cite{BSW2009} discovered that 
the center-of-mass (CM) energy can be arbitrarily high
if two particles which begin at rest at infinity collide near 
the horizon of a maximally rotating Kerr black hole~\cite{Kerr1963} and 
if the angular momentum of either particle is fine-tuned
to the critical value. 
They argue this scenario in the context of the collision of 
dark matter particles around intermediate-mass black holes.
This scenario is generalized to charged black holes~\cite{Zaslavskii2010_charged}, the Kerr-Newman family of black holes~\cite{WLGF2010} and 
general rotating black holes~\cite{Zaslavskii2010_rotating}.
A general explanation for the arbitrarily 
high CM energy is presented in terms of 
the Killing vectors and Killing horizon by 
Zaslavskii~\cite{Zaslavskii2010_explanation}.

The scenario by Ba\~nados, Silk and West~\cite{BSW2009} was subsequently 
criticized by several authors~\cite{Berti_etal2009,Jacobson_Sotiriou2010}.
One of the most important points is the 
limitations of the test particle approximation 
upon which their calculation relies. The validity of the test particle 
approximation is now under investigation. However, as we can see
for the exact analysis of the analogous system~\cite{Kimura_etal2011}, 
it is quite reasonable that the physical CM energy outside the horizon 
is bounded from above due to the violation of the test particle 
approximation. On the other hand, it is also reasonable that the 
upper limit on the CM energy is still 
considerably high in the situation where the test particle approximation
is good.

To circumvent the fine-tuning problem of the angular momentum, 
Harada and Kimura~\cite{Harada_Kimura2011} proposed a scenario, where 
the fine-tuning is naturally 
realized by the innermost stable circular orbit (ISCO) 
around a Kerr black hole~\cite{BPT1972}.
They discovered that the CM energy for the collision between an ISCO
particle and another generic particle becomes arbitrarily high 
in the limit of the maximal rotation of the black hole.
Even for the nonmaximally rotating black holes,
Grib and Pavlov~\cite{Grib_Pavlov2010_Kerr,Grib_Pavlov2010_particlecollisions} 
proposed a different scenario to obtain
the arbitrarily high CM energy of two colliding particles.
In this case, the 
particle with a near-critical angular momentum cannot reach the 
horizon from well outside through the geodesic motion
because of the potential barrier. 
In their scenario, the angular momentum of the particle
must be fine-tuned to the critical value 
through the preceding scattering near the horizon.

The geometry of a vacuum, stationary and asymptotically flat 
black hole is uniquely given by 
the Kerr metric~\cite{Kerr1963}. 
In the background of the Kerr spacetime, 
the expressions for the CM energy and its near-horizon limit 
are given for two colliding  
geodesic particles of the same rest mass, different 
energies and angular momenta in~\cite{Harada_Kimura2011}
and of different masses, energies and angular 
momenta in~\cite{Grib_Pavlov2010_particlecollisions},
although both are restricted to the motion on the equatorial plane.
It is quite important to extend the analysis to general geodesic particles 
not only because the analysis applies to 
realistic collisions in astrophysics
but also because we can get a deeper physical insight into 
the phenomenon itself. The general geodesic motion of 
massive and massless particles in the Kerr spacetime 
was analyzed by Carter~\cite{Carter1968}. See also~\cite{MTW1973,Poisson2004}.
The last stable orbit (LSO) is 
the counterpart of the ISCO for the nonequatorial motion
and defined by Sundararajan~\cite{Sundararajan2008}.

Based on Carter's formalism, we generalize the analysis of the CM energy of 
two colliding particles to 
general geodesic massive and massless particles.
In this paper, we adopt the test particle approximation and hence
neglect the effects of self-gravity and back reaction.
We then obtain an explicit expression for the CM energy  
of two colliding general geodesic particles at any spacetime point in 
the Kerr spacetime and 
derive a formula for the CM energy 
of two general geodesic particles colliding 
near the horizon of a Kerr black hole 
in terms of the conserved quantities of each particle and the 
polar angle.
We show that the collision with an arbitrarily high CM energy is possible
only in the limit to the horizon. We present the necessary and sufficient
condition to obtain an arbitrarily high CM energy and find that 
this condition is met only through the three scenarios, 
the direct collision scenario proposed 
by Ba\~nados, Silk and West~\cite{BSW2009}, 
the LSO (ISCO) collision scenario by Harada and Kimura~\cite{Harada_Kimura2011}
and the multiple scattering scenario by Grib and Pavlov~\cite{Grib_Pavlov2010_Kerr,Grib_Pavlov2010_particlecollisions}.
We find that the collision with an arbitrarily high CM energy is 
possible near the horizon of maximally rotating 
black holes not only at the equator but also at
the latitude up to 
$\mbox{acos}(\sqrt{3}-1)\simeq 42.94^{\circ}$
even if we do not 
admit the multiple scattering scenario.

This paper is organized as follows. 
In Sec. II, we briefly review general geodesic particles
in the Kerr spacetime. 
In Sec. III, we obtain an expression for the CM energy  
of two general geodesic particles at any spacetime point 
and then by taking the near-horizon limit 
obtain a general formula for the near-horizon collision. 
In Sec. IV, we classify critical particles, 
determine the region of the collision with 
an arbitrarily high CM energy with and without 
multiple scattering.
Section V is devoted to conclusion and discussion.
We use the units in which $c=G=1$ and 
the abstract index notation of Wald~\cite{Wald1984}.  

\section{General geodesic motion in the Kerr spacetime}
\subsection{The Kerr metric in the Boyer-Lindquist coordinates}
The line element in the Kerr spacetime 
in the Boyer-Lindquist coordinates is given by~\cite{Kerr1963,Wald1984,Poisson2004}   
\begin{eqnarray}
ds^{2}&=&-\left(1-\frac{2Mr}{\rho^{2}}\right)dt^{2}
-\frac{4Mar\sin^{2}\theta}{\rho^{2}}d\phi dt
+\frac{\rho^{2}}{\Delta}dr^{2}+\rho^{2}d\theta^{2} \nonumber \\
&&+\left(r^{2}+a^{2}+\frac{2Mra^{2}\sin^{2}\theta}{\rho^{2}}\right)
\sin^{2}\theta d\phi^{2} ,
\label{eq:Kerr_metric}
\end{eqnarray}
where $a$ and $M$ are the spin and mass parameters, respectively,
$\rho^{2}=r^{2}+a^{2}\cos^{2}\theta$ and $\Delta=r^{2}-2Mr+a^{2}$.
If $0\le a^{2}\le M^{2}$, 
$\Delta$ vanishes at $r=r_{\pm}=M\pm\sqrt{M^{2}-a^{2}}$, where 
$r=r_{+}$ and $r=r_{-}$ correspond to an event horizon and 
Cauchy horizon, respectively. Here, we denote $r_{+}=r_{H}$. 
In this coordinate system, the time 
translational and axial Killing vectors 
are given by
$\xi^{a}=(\partial/\partial t)^{a}$ and 
$\psi^{a}=(\partial/\partial \phi)^{a}$, 
respectively.
The surface gravity of the Kerr black hole is given by
$
\kappa=\sqrt{M^{2}-a^{2}}/(r_{H}^{2}+a^{2}).
$
Thus, the black hole has a vanishing surface gravity and hence is 
extremal for the maximal rotation $a^{2}=M^{2}$, while 
it is subextremal for the nonmaximal rotation $a^{2}<M^{2}$. 
The angular velocity of the 
horizon is given by 
\begin{equation}
\Omega_{H}=\frac{a}{r_{H}^{2}+a^{2}}.
\label{eq:Omega_H}
\end{equation}
The Killing vector $\chi^{a}=\xi^{a}+\Omega_{H}\psi^{a}$ is a null generator of the event horizon. 
We can assume $a\ge 0$ without loss of generality. 

\subsection{The Hamilton-Jacobi equation and the Carter constant}
We here briefly review general geodesic particles in the 
Kerr spacetime based on~\cite{MTW1973,Poisson2004}.
Let $S=S(\lambda,x^{\alpha})$ be the action as a function of 
the parameter $\lambda$ and coordinates $x^{\alpha}$, 
or the Hamilton-Jacobi function. The conjugate momentum is 
given by 
$p_{\alpha}=\partial S/\partial x^{\alpha}$.
Since the Hamiltonian for a geodesic particle is given by ${\cal H}=(1/2)\sum_{\mu,\nu}g^{\mu\nu}p_{\mu}p_{\nu}$,
we can explicitly write down the Hamilton-Jacobi equation with the metric~(\ref{eq:Kerr_metric}) in the following form:
\begin{eqnarray}
-\frac{\partial S}{\partial \lambda}&=&\frac{1}{2\rho^{2}}\left\{-\frac{1}{\Delta}\left[(r^{2}+a^{2})\frac{\partial S}{\partial t}+a \frac{\partial S}{\partial \phi}\right]^{2}
+\frac{1}{\sin^{2}\theta}\left[\frac{\partial S}{\partial \phi}
+a\sin^{2}\theta \frac{\partial S}{\partial t}\right]^{2}\right. \nonumber \\
&&+\left.\Delta\left(\frac{\partial S}{\partial r}\right)^{2}
+\left(\frac{\partial S}{\partial \theta}\right)^{2}\right\}.
\label{eq:HJ_explicit}
\end{eqnarray}
Since $\lambda$, $t$ and $\phi$ 
are cyclic coordinates, $S$ is written 
through the separation of variables as
\begin{equation}
S=\frac{1}{2}m ^{2}\lambda-Et+L\phi+S_{r}(r)+S_{\theta}(\theta),
\label{eq:variable_separation}
\end{equation}
where $m$, $E$ and $L$ are constants which correspond to the rest mass, 
conserved energy and angular momentum through
$m^{2}=-p_{a}p^{a}$, $E=-p_{t}=-\xi^{a}p_{a}$, and 
$L=p_{\phi}=\psi^{a}p_{a}$, respectively. 
Note that the proper time 
$\tau$ along the world line is given by $d\tau=md\lambda$
and the four velocity $u^{a}$ is given by $p^{a}=mu^{a}$
for a massive particle.

Substituting Eq.~(\ref{eq:variable_separation}) into 
Eq.~(\ref{eq:HJ_explicit}), 
we obtain
\begin{equation}
-\Delta \left(\frac{dS_{r}}{dr}\right)^{2}-m ^{2}r^{2}+\frac{[(r^{2}+a^{2})E-aL]^{2}}{\Delta}=\left(\frac{dS_{\theta}}{d\theta}\right)^{2}+m ^{2}a^{2}\cos^{2}\theta+\frac{1}{\sin^{2}\theta}[L-aE\sin^{2}\theta]^{2}.
\label{eq:Sr_Sth}
\end{equation}
It follows that both sides must be the same constant, 
which we denote with ${\cal K}$. That is to say, 
\begin{eqnarray}
{\cal K}&=&-\Delta \left(\frac{dS_{r}}{dr}\right)^{2}-m ^{2}r^{2}+\frac{[(r^{2}+a^{2})E-aL]^{2}}{\Delta}, 
\label{eq:K_r}\\
{\cal K}&=&\left(\frac{dS_{\theta}}{d\theta}\right)^{2}+m ^{2}a^{2}\cos^{2}\theta+\frac{1}{\sin^{2}\theta}[L-aE\sin^{2}\theta]^{2} 
.\label{eq:K_th}
\end{eqnarray}
Clearly, ${\cal K}\ge 0$ follows from Eq.~(\ref{eq:K_th}). 
The Carter constant ${\cal Q}$ is a conserved quantity defined by
${\cal Q}\equiv {\cal K}-(L-aE)^{2}$ or 
\begin{eqnarray}
{\cal Q}=\left(\frac{dS_{\theta}}{d\theta}\right)^{2}+\cos^{2}\theta\left[a^{2}(m ^{2}-E^{2})+\frac{L^{2}}{\sin^{2}\theta}\right].
\label{eq:Carter_constant}
\end{eqnarray}
Note that ${\cal Q}$ can be negative but ${\cal Q}+(L-aE)^{2}\ge 0$ must be satisfied. On the other hand, 
we find ${\cal Q}\ge 0$ if $m^{2}\ge E^{2}$ from 
Eq.~(\ref{eq:Carter_constant}).

We integrate Eqs.~(\ref{eq:K_r}) and (\ref{eq:K_th}) to give
\begin{equation*}
S_{\theta}=\sigma_{\theta}\int^{\theta}d\theta\sqrt{\Theta}, \quad
S_{r}=\sigma_{r}\int^{r}dr\frac{\sqrt{R}}{\Delta},
\end{equation*}
where the choices of the two signs $\sigma_{\theta}=\pm 1$
and $\sigma_{r}=\pm 1$ are independent and
\begin{eqnarray}
\Theta&=&\Theta(\theta)={\cal Q}-\cos^{2}\theta\left[a^{2}(m ^{2}-E^{2})+\frac{L^{2}}{\sin^{2}\theta}\right], 
\label{eq:Theta}\\
R&=& R(r)=P(r)^{2}-\Delta(r) [m ^{2}r^{2}+(L-aE)^{2}+{\cal Q}], 
\label{eq:R}\\
P&=&P(r)= (r^{2}+a^{2})E-aL.
\label{eq:Theta_R_P}
\end{eqnarray}
Thus, we obtain the Hamilton-Jacobi function. 
Note that for the allowed motion both 
$\Theta\ge 0$ and $R\ge 0$ must be satisfied.  

Using
$dx^{\alpha}/d\lambda=p^{\alpha}=\sum_{\beta}g^{\alpha\beta}p_{\beta}$,
we obtain
\begin{eqnarray}
\rho^{2}\frac{dt}{d\lambda}&=&-a(aE\sin^{2}\theta-L)+\frac{(r^{2}+a^{2})P}{\Delta}, 
\label{eq:dtdlambda}\\
\rho^{2}\frac{dr}{d\lambda}
&=&\sigma_{r} \sqrt{R}, \label{eq:drdlambda}\\
\rho^{2}\frac{d\theta}{d\lambda}&=& \sigma_{\theta}\sqrt{\Theta}, 
\label{eq:dthetadlambda}\\
\rho^{2}\frac{d\phi}{d\lambda}&=& -\left(aE-\frac{L}{\sin^{2}\theta}\right)+\frac{aP}{\Delta}.
\end{eqnarray}

\subsection{Properties of geodesic particles in the Kerr spacetime}
From Eqs.~(\ref{eq:Theta}) and (\ref{eq:dthetadlambda}), we can see
${\cal Q}=0$ must be satisfied 
for a particle moving on the equatorial plane $\theta=\pi/2$. As we can see in Eqs.~(\ref{eq:Theta}) and (\ref{eq:dthetadlambda}),  if $L\ne 0$, the particle 
oscillates with respect to $\theta$ and 
never reaches the rotational axis $\theta=0$ or $\pi$.
A special treatment is needed for a particle which crosses the 
rotational axis $\theta=0$ or $\pi$, which is a coordinate singularity. 
To have a regular limit to the axis in Eq.~(\ref{eq:Sr_Sth}), 
we impose $L=0$ to such a particle.
Only particles with $L=0$ can cross the rotational axis.

Equation~(\ref{eq:drdlambda}) imply 
\begin{equation}
\frac{1}{2}\left(\frac{dr}{d\lambda}\right)^{2}+\frac{r^{4}}{\rho^{4}}V_{\rm eff}(r)=0,
\label{eq:mechanical_energy}
\end{equation}
where
\begin{equation}
V_{\rm eff}(r)\equiv -\frac{R(r)}{2r^{4}}.
\label{eq:effective_potential}
\end{equation}
Since $r^{4}/\rho^{4}$ is nonzero and finite outside the horizon, 
$V_{\rm eff}$ plays a role similar to the effective potential for the motion on 
the equatorial plane, although there is a coupling with $\theta$ 
in Eq.~(\ref{eq:mechanical_energy}). The allowed and prohibited regions are given by 
$V_{\rm eff}(r)\le 0$ and $V_{\rm eff}(r)>0$, respectively. 
Since $V_{\rm eff}(r)\to  (m^{2}-E^{2})/2$ 
as $r\to \infty$, the sign of $(m^{2}-E^{2})$ governs the particle motion 
far away from the black hole. 
A particle is bound, marginally bound 
and unbound if $m^{2}>E^{2}$, $m^{2}=E^{2}$ and $m^{2}<E^{2}$, respectively.

Let us
consider special null geodesics with ${\cal K}=0$. 
Then, $L=aE\sin^{2}\theta$, 
${\cal Q}=-(L-aE)^{2}=-(aE\cos^{2}\theta)^{2}$ and hence 
$\Theta=0$. Thus, $\theta=$const, $P=E\rho^{2}$ and $R=E^{2}\rho^{4}$.
Then, we obtain simple geodesics:
\begin{equation*}
\frac{dt}{d\lambda}= \frac{E(r^{2}+a^{2})}{\Delta},\quad
\frac{dr}{d\lambda}=\sigma_{r} E,\quad
\frac{d\theta}{d\lambda}=0, \quad
\frac{d\phi}{d\lambda}=\frac{aE}{\Delta}.
\end{equation*}
This means that for any value of $\theta$, 
there are always ingoing and outgoing null geodesics along which $\theta=$const.
These geodesics are called outgoing (ingoing) 
principal null geodesics for $\sigma_{r}=1$ ($-1$).

Since we are considering causal geodesics parametrized from 
the past to the future, we need to impose $dt/d\lambda\ge 0$ 
along the geodesic. This is called the ``forward-in-time'' condition.
In particular, 
as seen from Eq.~(\ref{eq:dtdlambda}), 
this condition reduces to 
\begin{equation}
E - \Omega_{H}L\ge 0,
\label{eq:forward_in_time_horizon}
\end{equation} 
in the near-horizon limit, 
where we have used Eq.~(\ref{eq:Omega_H}).
Shortly, the angular momentum must be smaller than the critical value 
$L_{c}\equiv \Omega_{H}^{-1}E$.
This condition is identical to the forward-in-time condition 
near the horizon for particles restricted on the equatorial plane.
We refer to particles with the angular momentum $L=L_{c}$, $L<L_{c}$
and $L>L_{c}$ as critical, subcritical and supercritical particles,
respectively. 
We can easily see that $L\le L_{c}$ is equivalent to the condition 
\begin{equation*}
-\chi^{a}p_{a}\ge 0,
\end{equation*}
for the horizon-generating Killing vector $\chi^{a}$ and the four 
momentum $p_{a}$ of the particle. This must clearly hold near the horizon for the subextremal 
black hole because $\chi^{a}$ is future-pointing timelike there and $p^{a}$ 
is future-pointing timelike or null.

\section{CM energy of two colliding general geodesic particles}

\subsection{\label{sec:CM_energy_different_mass}
CM energy of two colliding particles of different rest masses}

Let particles 1 and 2 of rest masses $m_{1}$ and $m_{2}$ have four momenta 
$p^{a}_{1}$ and $p^{a}_{2}$, respectively.
The sum of the two momenta is given by
\begin{equation*}
p_{\rm tot}^{a}=p_{1}^{a}+p_{2}^{a}.
\end{equation*}
The CM energy $E_{\rm cm}$ of the two particles is then given by
\begin{equation}
E_{\rm cm}^{2}=-p^{a}_{\rm tot}p_{{\rm tot}a}=m_{1}^{2}+m_{2}^{2}
-2g^{ab}p_{1a}p_{2b}.
\label{eq:center-of-mass_energy_different_mass}
\end{equation}

Clearly, this applies for both massive and massless particles.
Since $E_{\rm cm}$ is a scalar, it does not depend on the coordinate 
choice in which we evaluate it. This is the reason why we can safely determine
the CM energy in the Boyer-Lindquist coordinates in spite of the coordinate
singularity on the horizon.

\subsection{CM energy of two colliding 
particles in the Kerr spacetime}

As seen in Sec.~\ref{sec:CM_energy_different_mass}, the CM energy of two particles is determined by calculating $-g^{ab}p_{1a}p_{2b}$.
Using Eq.~(\ref{eq:Kerr_metric}), 
the CM energy is then calculated to give
\begin{eqnarray}
E_{\rm cm}^{2}&=&m_{1}^{2}+m_{2}^{2}+\frac{2}{\rho^{2}}\left[\frac{P_{1}P_{2}-\sigma_{1r}\sqrt{R_{1}}\sigma_{2r}\sqrt{R_{2}}}{\Delta}-\frac{(L_{1} -a\sin^{2}\theta E_{1})(L_{2}-a\sin^{2}\theta E_{2})}{\sin^{2}\theta}\right. \nonumber \\
&& \left. -\sigma_{1\theta}\sqrt{\Theta_{1}}\sigma_{2\theta}\sqrt{\Theta_{2}}\right],
\label{eq:E_cm_explicit}
\end{eqnarray}
where and hereafter $E_{i}$, $L_{i}$, ${\cal Q}_{i}$, ${\cal K}_{i}$, 
$P_{i}=P_{i}(r)$, $R_{i}=R_{i}(r)$ and $\Theta_{i}=\Theta_{i}(\theta)$
are $E$, $L$, ${\cal Q}$, ${\cal K}$, 
$P=P(r)$, $R=R(r)$ and $\Theta=\Theta(\theta)$ for particle $i$, respectively.
This is surprisingly simple in spite of the generality of this expression.
This is due to the separability of the Hamilton-Jacobi equation in the 
Kerr spacetime.
From Eqs.~(\ref{eq:K_th}), (\ref{eq:Theta}) with $\Theta\ge 0$
and $\rho^{2}=r^{2}+a^{2}\cos^{2}\theta$, it follows that
\begin{eqnarray*}
\left|\frac{L-a\sin^{2}\theta E}{\sin\theta}\right|\le \sqrt{{\cal K}}, \quad 
\sqrt{\Theta}
\le \sqrt{|{\cal Q}|+a^{2}|m^{2}-E^{2}|}, \quad
r_{H}^{2}\le \rho^{2}\le r^{2}+a^{2}
\end{eqnarray*}
outside the horizon.
Moreover, in the limit $r\to \infty$, we obtain
\begin{eqnarray*}
E_{\rm cm}^{2}\to m_{1}^{2}+m_{2}^{2}+
2\left(E_{1}E_{2}-\sigma_{1r}\sqrt{E_{1}^{2}-m_{1}^{2}}\sigma_{2r}\sqrt{E_{2}^{2}-m_{2}^{2}}\right).
\end{eqnarray*}
Therefore, Eq.~(\ref{eq:E_cm_explicit}) assures that 
if all conserved quantities $m_{i}$, $E_{i}$, $L_{i}$, 
${\cal K}_{i}$ are bounded from above, $E_{\rm cm}$
is also bounded from above except in
the limit to the horizon where $\Delta= 0$.
In other words, only if the collision occurs near the horizon,
the CM energy can be unboundedly high.

\subsection{CM energy of two particles colliding near the horizon}
\label{subsec:CM_energy_near-horizon_collision}
If $\sigma_{1r}$ and $\sigma_{2r}$
have different signs near the horizon, the CM energy 
for two colliding particles 
necessarily diverges in the near-horizon limit $\Delta \to 0$ as 
\begin{equation*}
E_{\rm cm}^{2}\approx 4\frac{(r_{H}^{2}+a^{2})^{2}}
{r_{H}^{2}+a^{2}\cos^{2}\theta}\frac{(E_{1}-\Omega_{H}L_{1})(E_{2}-\Omega_{H}L_{2})}{\Delta}, 
\end{equation*}
where both particles are assumed to be subcritical.
However, $\sigma_{1r}$ and $\sigma_{2r}$ must not have 
different signs right on the black hole horizon.
 
Then, we assume that $\sigma_{1r}$ and $\sigma_{2r}$ have the same sign.
In the near-horizon limit $r\to r_{H}$, we can see that
$(P_{1}P_{2}-\sqrt{R_{1}}\sqrt{R_{2}})$ vanishes.
In fact, it is easy to show 
\begin{eqnarray*}
&&\lim_{r\to r_{H}}\frac{P_{1}P_{2}-\sqrt{R_{1}}\sqrt{R_{2}}}{\Delta}
\nonumber \\
&=& \frac{m _{1}^{2}r_{H}^{2}+{\cal K}_{1}}{2}\frac{(r_{H}^{2}+a^{2})E_{2}-aL_{2}}{(r_{H}^{2}+a^{2})E_{1}-aL_{1}}+\frac{m _{2}^{2}r_{H}^{2}+{\cal K}_{2}}{2}\frac{(r_{H}^{2}+a^{2})E_{1}-aL_{1}}{(r_{H}^{2}+a^{2})E_{2}-aL_{2}},
\end{eqnarray*}
where we have assumed subcritical particles.
Therefore, the CM energy of two general geodesic particles in the near-horizon limit is written as
\begin{eqnarray}
E_{\rm cm}^{2}&=&m_{1}^{2}+m_{2}^{2}+
\frac{1}{r_{H}^{2}+a^{2}\cos^{2}\theta}
\left[(m_{1}^{2}r_{H}^{2}+{\cal K}_{1})\frac{E_{2}-\Omega_{H}L_{2}}{E_{1}-\Omega_{H}L_{1}}+(m_{2}^{2}r_{H}^{2}+{\cal K}_{2})\frac{E_{1}-\Omega_{H}L_{1}}{E_{2}-\Omega_{H}L_{2}} \right. \nonumber \\
&& \left. -\frac{2(L_{1}-a\sin^{2}\theta E_{1})(L_{2}-a\sin^{2}\theta E_{2})}{\sin^{2}\theta}
-2\sigma_{1\theta}\sqrt{\Theta_{1}}\sigma_{2\theta}\sqrt{\Theta_{2}}\right],
\label{eq:general_formula}
\end{eqnarray}
where we have used Eq.~(\ref{eq:Omega_H}).
We can now find that the necessary and sufficient condition 
to obtain an arbitrarily high CM energy is that 
\begin{equation*}
(m_{1}^{2}r_{H}^{2}+{\cal K}_{1})\frac{E_{2}-\Omega_{H}L_{2}}{E_{1}-\Omega_{H}L_{1}}+(m_{2}^{2}r_{H}^{2}+{\cal K}_{2})\frac{E_{1}-\Omega_{H}L_{1}}{E_{2}-\Omega_{H}L_{2}}
\end{equation*}
is arbitrarily large.
It is also clear that the necessary condition for the CM energy to be unboundedly high 
is that $(E-\Omega_{H} L)$ is arbitrarily close to zero for either of the 
two particles. That is to say, either of the two particles must be 
arbitrarily near-critical. 

Furthermore, we can show $(m^{2}r_{H}^{2}+{\cal K})$ is 
bounded from below by a positive value for critical particles with $E\ne 0$. This is trivial for massive particles. 
For the massless case, from Eq.~(\ref{eq:K_th}), 
we find for the critical particle
\begin{equation*}
{\cal K}\ge \frac{[\Omega_{H}^{-1}E-aE\sin^{2}\theta ]^{2}}{\sin^{2}\theta}
=   \left(\frac{r_{H}^{2}+a^{2}\cos^{2}\theta}{a\sin\theta}\right)^{2}E^{2}
\ge  \left(\frac{r_{H}^{2}}{a}\right)^{2}E^{2}>0,
\end{equation*}
where we have used Eq.~(\ref{eq:Omega_H}). Therefore, 
$(m^{2}r_{H}^{2}+{\cal K})$ is bounded from below by a positive value 
except for the case where $m=E=L=0$. 
Although this exceptional case might be physically 
meaningful, we do not need to deal with it for the present purpose.
Note that since a null geodesic is principal null if and only if 
${\cal K}=0$,  
no principal null geodesic can be critical 
as a contraposition. In fact, any principal null geodesic
turns out to be subcritical because $L=aE\sin^{2}\theta\le a E 
<\Omega_{H}^{-1}E=L_{c}$.

Unless the critical particle is massless with vanishing energy, 
the necessary and sufficient condition 
to obtain an arbitrarily high CM energy reduces so that the ratio
\begin{equation*}
\frac{E_{1}-\Omega_{H} L_{1}}{E_{2}-\Omega_{H}L_{2}}
\end{equation*}
is arbitrarily large or arbitrarily close to zero.
If this ratio is arbitrarily close to zero, 
Eq.~(\ref{eq:general_formula}) is approximated as
\begin{equation*}
E_{\rm cm}^{2}\approx 
\frac{m_{1}^{2}r_{H}^{2}+{\cal K}_{1}}
{r_{H}^{2}+a^{2}\cos^{2}\theta}
\frac{E_{2}-\Omega_{H}L_{2}}{E_{1}-\Omega_{H}L_{1}}.
\end{equation*}

For the particles
moving on the equatorial plane, we set $\theta=\pi/2$ and 
${\cal Q}=0$. Then, Eq.~(\ref{eq:general_formula}) reduces to
\begin{eqnarray}
E_{\rm cm}^{2}&=&m_{1}^{2}+m_{2}^{2}+\frac{1}{r_{H}^{2}}
\left\{\left[m_{1}^{2}r_{H}^{2}+(L_{1}-aE_{1})^{2}\right]\frac{E_{2}-\Omega_{H}L_{2}}{E_{1}-\Omega_{H}L_{1}}\right.\nonumber \\
&& \left. +\left[m_{2}^{2}r_{H}^{2}+(L_{2}-aE_{2})^{2}\right]\frac{E_{1}-\Omega_{H}L_{1}}{E_{2}-\Omega_{H}L_{2}}-2(L_{1}-a E_{1})(L_{2}-a E_{2})\right\}.
\label{eq:Harada_Kimura_variant}
\end{eqnarray}
If we further assume that the colliding particles have the same nonzero 
rest mass $m_{0}$, it is easy to explicitly confirm that Eq.~(\ref{eq:Harada_Kimura_variant}) 
coincides with the formula~(3.5) of Harada and Kimura~\cite{Harada_Kimura2011} or
\begin{equation*}
\frac{E_{\rm cm}}{2m_{0}}=
\sqrt{1+\frac{4M^{2} m_{0}^{2} [(E_{1}-\Omega_{H}L_{1})-(E_{2}-\Omega_{H}L_{2})]^{2}
+(E_{1}L_{2}-E_{2}L_{1})^{2}}{16M^{2} m_{0}^{2} (E_{1}-\Omega_{H}L_{1})(E_{2}-\Omega_{H}L_{2})}}
\end{equation*}
in the present notation.

\section{Collision with an arbitrarily high CM energy}

\subsection{\label{subsec:classification}
Classification of critical particles}

Since either of the two colliding particles must be arbitrarily
near-critical to obtain an arbitrarily high CM energy, 
we here study critical particles, i.e. particles with 
the critical angular momentum $L=L_{c}\equiv\Omega_{H}^{-1}E$.
Although the critical particle 
may be prohibited to reach the horizon
or it can do so only 
after an infinite proper time, the critical particle 
still characterizes near-critical particles 
as a {\it limit critical particle}.

From Eq.~(\ref{eq:R}), we find 
\begin{equation*}
R(r_{H})=(r_{H}^{2}+a^{2})^{2}(E-\Omega_{H}L)^{2}.
\end{equation*}
Therefore, $R(r_{H})\ge 0$. In particular, only for critical particles, i.e.
$E-\Omega_{H}L=0$, $R(r_{H})=0$ holds. For the first derivative, 
from Eq.~(\ref{eq:R}), we find 
\begin{equation*}
R'(r_{H})=4r_{H}(r_{H}^{2}+a^{2})E(E-\Omega_{H}L)-2(r_{H}-M)(m ^{2}r_{H}^{2}+{\cal K}).
\end{equation*}
As we have seen in 
Sec.~\ref{subsec:CM_energy_near-horizon_collision}, 
the factor $(m^2 r_H^2 + {\cal K})$ 
is positive for critical particles.
Therefore, we conclude $R'(r_{H})\le 0$ for the critical particle
because $r_{H}\ge M$.

If $R'(r_{H})=0$ for the critical particle, 
the Kerr black hole is necessarily extremal. 
In this case, $R$ for the critical particle becomes
\begin{equation}
R=(r-M)^{2}[(E^{2}-m ^{2})r^{2}+2ME^{2}r-{\cal Q}]
\label{eq:R_critical_extremal}
\end{equation}
and hence 
\begin{equation}
R''(r_{H})=2\left[(3E^{2}-m^{2})M^{2}-{\cal Q}\right].
\end{equation}
Although one might expect a circular orbit
of massive particles on the horizon for $R=R'=0$ there, this is 
fake as is proven in~\cite{Harada_Kimura2011}.

Suppose $R'(r_{H})=0$ and 
$R''(r_{H})>0$, i.e. $(3E^{2}-m^{2})M^{2}>{\cal Q}$. Then, 
$R(r)> 0$ at least 
in the vicinity of the horizon for the critical particle. 
This class includes what Ba\~nados, Silk and West~\cite{BSW2009} 
originally assume 
and we refer to this class as class I.
A critical particle of class I can reach the horizon along 
a geodesic from outside after an infinite proper time.

The condition $R'(r_{H})=0$ and $R''(r_{H})=0$, i.e. 
$(3E^{2}-m^{2})M^{2}={\cal Q}$ 
corresponds to the marginal case 
and this is exactly the situation
studied in Harada and Kimura~\cite{Harada_Kimura2011} for the equatorial case.
This is of particular physical interest because 
the sequence of the prograde ISCO particle
converges to this limit, where the fine-tuning of the angular momentum 
is naturally realized in the astrophysical context.
Since the function $R$ takes an inflection point 
at the ISCO radius and hence $R=R'=R''=0$ 
there, the potential of the limit critical particle should 
satisfy $R(r_{H})=R'(r_{H})=R''(r_{H})=0$.
Hence, we treat this class as a separate case 
and refer to this class as class II.
For ${\cal Q}\ne 0$, the critical particle of this class
corresponds to the limit critical particle of the sequence of 
particles orbiting 
the inclined LSO in the limit $a\to 1$ 
according to the definition 
$R=R'=R''=0$ given by Sundararajan~\cite{Sundararajan2008}.
This means that the scenario of the high-velocity 
collision of an ISCO particle generalizes to the nonequatorial case,
as the high-velocity collision of an LSO particle.

We can also consider the case where $R'(r_{H})=0$ and 
$R''(r_{H})<0$, i.e. $(3E^{2}-m^{2})M^{2}<{\cal Q}$. 
Although this case has not been mentioned so far in the literature 
in the present context, 
we refer to this class as class III.  
The behavior of the critical particles of this class 
is similar to that of the critical particles of class IV 
described below.

The possibility $R'(r_{H})< 0$ for the limit critical particle
was first raised by Grib and Pavlov~\cite{Grib_Pavlov2010_Kerr,Grib_Pavlov2010_particlecollisions}.
This is possible only for the subextremal black hole.
We refer to this class as class IV.
In the sequence approaching the critical particle of this class,
near-critical particles with an angular momentum 
$L=L_{c}-\delta$ for sufficiently small $\delta(>0)$ can 
approach the horizon along a geodesic only 
from the vicinity of the horizon.
Such near-critical particles are possible only 
through multiple scattering because they 
must be inside the potential barrier before the relevant collision.
All the critical particles in a subextremal black hole
belong to this class.

In principle, one might expect that there is 
a critical particle with $R'(r_{H})>0$.
Such particles should have similar characteristics to those of class I.
However, as we have seen, such a critical particle does not exist 
in the Kerr spacetime.

\begin{figure}[htbp]
\begin{center}
\includegraphics[width=0.8\textwidth]{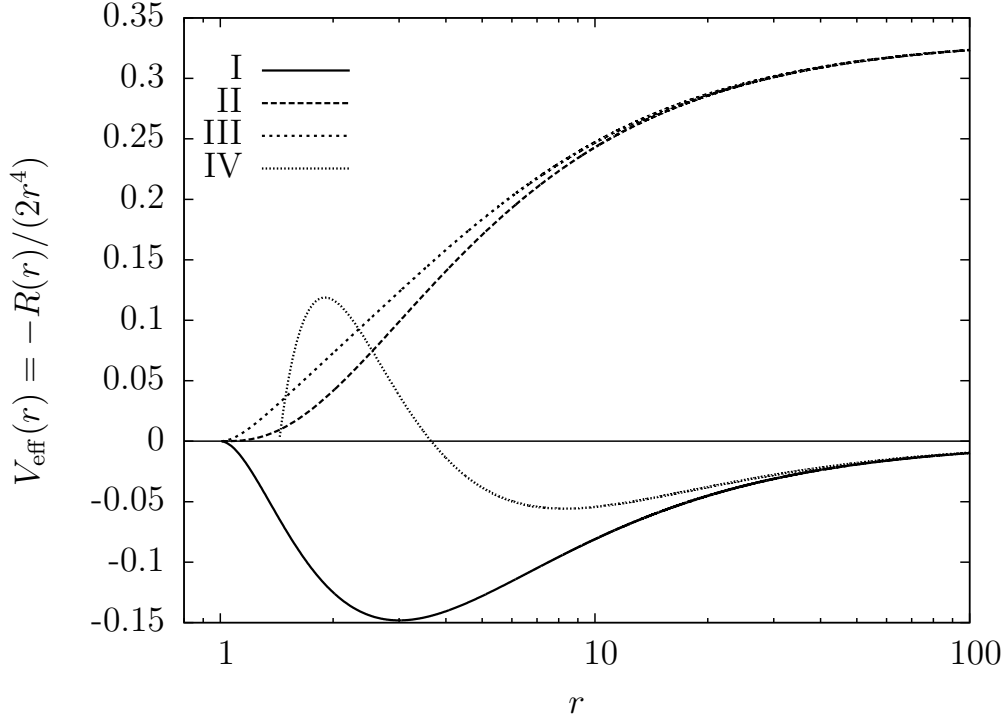}
\caption{\label{fg:processes}
The examples of the effective potential $V_{\rm eff}(r)=-R(r)/(2r^{4})$ 
for the critical particles. The solid, long-dashed, dashed 
and short-dashed curves
show the potentials for the particles of classes 
I ($M=a=1$, $m=E=1$, $L=2$, ${\cal Q}=0$),
II ($M=a=1$, $m=1$, $E=1/\sqrt{3}$, 
$L=2/\sqrt{3}$, ${\cal Q}=0$),
III
($M=a=1$, $E=1/\sqrt{3}$, $L=2/\sqrt{3}$, ${\cal Q}=1$), 
and IV ($M=1$, $a=0.9$, $m=E=1$, $L=2$, ${\cal Q}=0$),
respectively.
}
\end{center}
\end{figure}

The conditions for $R(r_{H})$, $R'(r_{H})$ and $R''(r_{h})$
are easily converted to those for $V_{\rm eff}(r_{H})$, $V_{\rm eff}'(r_{H})$
and $V_{\rm eff}''(r_{H})$ in terms of the effective potential $V_{\rm eff}(r)$
defined by Eq.~(\ref{eq:effective_potential}).
Table~\ref{tb:critical_particles} summarizes the four classes of 
critical particles and the three scenarios of the collision with 
an arbitrarily high CM energy. 
Note that classes III and IV belong to the same scenario so that
we have four classes in spite of three scenarios.
Figure~\ref{fg:processes} shows the examples of the effective 
potentials for the critical particles of these four classes.
Although the classification in this subsection only concerns 
the signs of the function $R$ and 
its derivatives at the horizon, it turns out that 
critical particles of class I with $E^{2}\ge m^{2}$ 
correspond to the direct collision scenario from infinity,
as we will see in Sec.~\ref{subsec:direct_collision_from_infinity}.

\begin{table}[htb]
\begin{center}
\caption{Classification of critical particles and the collision scenarios}
\label{tb:critical_particles} 
\begin{tabular}{|c|c|c|c|c|}
\hline
\multicolumn{1}{|c|} {Class} & $R(r)$ at $r=r_{H}$ & Scenario & Reference &  Parameter region \\
\hline\hline
I & $R=R'=0$, $R''>0$ & direct collision & \cite{BSW2009}& $a^{2}=M^{2}$, $3E^{2}> m^{2}$, ${\cal Q}<(3E^{2}-m^{2})M^{2}$ \\
\hline
II & $R=R'=R''=0$ & LSO collision &\cite{Harada_Kimura2011} &  
$a^{2}=M^{2}$, $3E^{2}\ge m^{2}$, ${\cal Q}=(3E^{2}-m^{2})M^{2}$ \\
\hline
III & $R=R'=0$, $R''<0$ & multiple scattering & --& 
$a^{2}=M^{2}$, ${\cal Q}>(3E^{2}-m^{2})M^{2}$\\
\hline
IV& $R=0$, $R'<0$ & multiple scattering  & \cite{Grib_Pavlov2010_Kerr,Grib_Pavlov2010_particlecollisions}& $0< a^{2}<M^{2}$  \\
\hline
\end{tabular}
\end{center}
\end{table}

\subsection{\label{subsec:belt}
The high-velocity collision belts on the extremal Kerr black hole}

It is not necessarily clear how the fine-tuning of the angular 
momentum is realized near the horizon through multiple scattering
processes. Hence, hereafter we 
concentrate on the direct collision scenario and 
the LSO collision scenario. Then, 
critical particles of classes I and II are relevant, which
are possible only if the black hole is extremal and 
\begin{equation}
(3E^{2}-m^{2})M^{2}\ge {\cal Q}
\label{eq:upperboundonQ}
\end{equation}
is satisfied for the critical particle, as 
we have seen in the previous section.
Together with Eq.~(\ref{eq:Carter_constant}),
${\cal Q}$ must satisfy the following condition:
\begin{equation}
\cos^{2}\theta\left[M^{2}(m ^{2}-E^{2})+\frac{4M^{2}E^{2}}{\sin^{2}\theta}\right]
\le  {\cal Q} \le (3E^{2}-m^{2})M^{2},
\label{eq:range_of_Q}
\end{equation} 
where $a^{2}=M^{2}$ and $L=L_{c}=2ME$ have been used.
We will see here whether this condition restricts the polar angle.
From Eq.~(\ref{eq:range_of_Q}), the following condition must be satisfied:
\begin{equation}
(m ^{2}-E^{2})\sin^{4}\theta+2(4E^{2}-m ^{2})\sin^{2}\theta-4E^{2}\ge 0.
\label{eq:inequality}
\end{equation}
Conversely, if Eq.~(\ref{eq:inequality}) holds, we can always find ${\cal Q}$
which satisfies Eq.~(\ref{eq:range_of_Q}).

For the marginally bound orbit $m ^{2}= E^{2}$, we can easily find
from Eq.~(\ref{eq:inequality}) 
\begin{equation*}
\sin\theta\ge \sqrt{\frac{2}{3}}.
\end{equation*}
This means that critical particles 
can occur only on the belt 
between latitudes $(\pi/2-\theta)=\pm \mbox{acos}\sqrt{2/3}\simeq \pm 35.26^{\circ}$. 
It is also easy to generalize this bound to nonmarginally bound particles
because the left-hand side of inequality (\ref{eq:inequality}) is only 
quadratic with respect to $\sin^{2}\theta$. 
The result is that $E^{2}$ must satisfy $3E^{2}\ge m ^{2}$ and then 
$\theta$ must satisfy the following condition:
\begin{equation}
\sin\theta\ge \sqrt{\frac{-(4E^{2}-m ^{2})+\sqrt{12E^{4}-4E^{2}m ^{2}+m ^{4}}}
{m ^{2}-E^{2}}}.
\label{eq:belt_general}
\end{equation}
Therefore, the absolute value of the 
latitude must be lower than the angle $\alpha (E,m)$, where 
\begin{equation*}
\alpha(E,m)=\mbox{acos}\left(
\sqrt{\frac{-(4E^{2}-m ^{2})+\sqrt{12E^{4}-4E^{2}m ^{2}+m ^{4}}}
{m ^{2}-E^{2}}}\right).
\end{equation*}
The above applies to both bound ($m ^{2}>E^{2}$) and unbound ($m ^{2}<E^{2}$)
particles.

\begin{figure}
\includegraphics[width=0.8\textwidth]{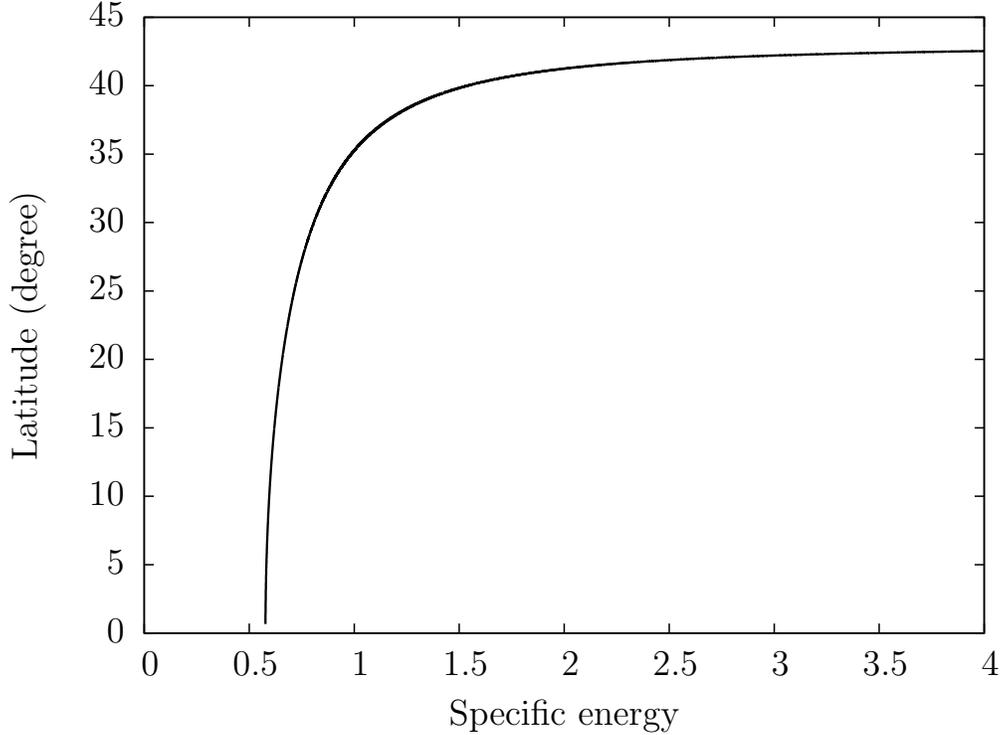}
\caption{\label{fg:latitude}
The highest absolute value of 
the latitude for the critical particles of classes I and II to occur on the 
extremal Kerr black hole as a function of the specific energy.
}
\end{figure}

For $3E^{2}=m ^{2}$, 
Eqs.~(\ref{eq:range_of_Q}) and (\ref{eq:inequality}) imply 
$\theta=\pi/2$ and ${\cal Q}=0$ so that  
the critical particle belongs to class II 
and it is on the equatorial plane.
This is an ISCO particle for the maximal black hole spin.
In the limit $E^{2}\to m ^{2}$, the right-hand side of 
Eq.~(\ref{eq:belt_general}) approaches $\sqrt{2/3}$ and hence reproduces the 
result for the marginally bound particles. It is quite intriguing to see
the limit $E^{2}\to \infty$. In this limit, the right-hand side 
of Eq.~(\ref{eq:belt_general}) approaches $\sqrt{3}-1$ and hence
\begin{equation*}
\sin\theta\ge \sqrt{3}-1.
\end{equation*}
Noting that the right-hand side of Eq.~(\ref{eq:belt_general}) is 
monotonically decreasing as a function of $E^{2}$, 
the belt where critical 
particles can occur becomes larger as the 
energy of the particle is greater. However, the 
latitude limit of the belt does not reach the poles but approaches
$\pm \mbox{acos}(\sqrt{3}-1)\simeq \pm 42.94^{\circ}$ as 
the energy of the particle is increased to infinity.
In other words, no critical particle occurs
with the latitude higher than this angle.
The highest absolute value of the latitude is shown 
in Fig.~\ref{fg:latitude} as a function of the specific energy
of the particle.

For a massless particle, i.e. $m=0$, 
Eq.~(\ref{eq:belt_general}) simply reduces to 
\begin{equation*}
\sin\theta\ge \sqrt{3}-1,
\end{equation*}
irrespective of the energy of the particle. Thus, the highest absolute value of the latitude is 
$\mbox{acos}(\sqrt{3}-1)\simeq 42.94^{\circ}$ if the near-critical 
particle is massless.

\begin{figure}
\includegraphics[width=0.5\textwidth]{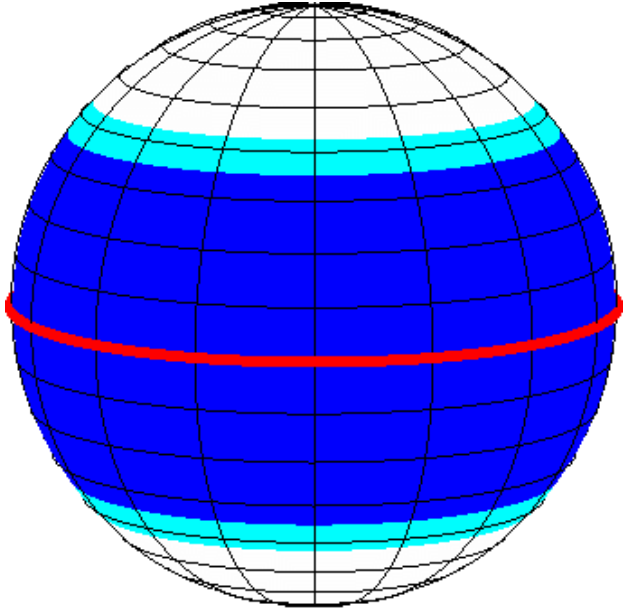}
\caption{\label{fg:high-velocity_collision_belt}
The belts of high-velocity collision on the extremal Kerr black hole. 
The red (solid thick) line shows the equator. 
The collisions with an arbitrarily high CM energy occur
on the belt colored with blue and cyan (shaded darkly and lightly)
between latitudes $\pm \mbox{acos}(\sqrt{3}-1)\simeq \pm 42.94^{\circ}$
if we allow all the critical particles.
On the other hand, such collisions occur on the belt colored with blue
(shaded darkly) 
between latitudes $\pm \mbox{acos}\sqrt{2/3}\simeq \pm 35.26^{\circ}$
if we only allow bound and marginally bound massive 
critical particles.
On the uncolored (unshaded) region, the collision with 
an arbitrarily high CM energy is prohibited.}
\end{figure}

The result is schematically shown in 
Fig.~\ref{fg:high-velocity_collision_belt}. 
This figure shows the regions of high-velocity collision on the extremal Kerr
black hole. The red (solid thick) line shows the equator.
The collisions with an arbitrarily high CM energy occur
on the belt colored with blue and cyan (shaded darkly and lightly)
if we allow all the critical particles.
On the other hand, such collisions occur on the belt
colored with blue (shaded darkly)
if we only allow bound and marginally bound massive 
critical particles.
On the uncolored (unshaded) region, the collision with 
an arbitrarily high CM energy is prohibited.

\subsection{
Direct collision from infinity with nonequatorial geodesics}
\label{subsec:direct_collision_from_infinity}
Ba\~nados, Silk and West~\cite{BSW2009} originally proposed a scenario where 
a massive particle which is at rest at infinity, i.e. $E^{2}=m^{2}$,  
with a near-critical angular momentum $L\approx L_{c}=2Mm$
falls towards an extremal Kerr black hole on the equatorial plane   
and collides with another particle near the horizon with an 
arbitrarily high CM energy in the limit $L\to L_{c}$.

First, we only relax the restriction of 
the equatorial motion in their scenario 
and see whether the CM energy can still be arbitrarily high.
In the original scenario by Ba\~nados, Silk and West~\cite{BSW2009}, 
it is important that
the geodesic motion from infinity to the horizon is allowed. 
This means that the function $R(r)$ must be positive 
for $r_{H}<r<\infty$, with which we have not been concerned 
in Secs.~\ref{subsec:classification} and \ref{subsec:belt}.
As seen in Eq.~(\ref{eq:R_critical_extremal})
with $E^{2}=m^{2}$, 
this is the case if and only if ${\cal Q}\le 2 m^{2}M^{2}$. 
Then, marginally bound particles with a near-critical 
angular momentum $L=L_{c}-\delta$ for a sufficiently small $\delta(>0)$ 
can approach the horizon from infinity 
and collide with another particle near the horizon.
Actually, the condition ${\cal Q}\le 2 m^{2}M^{2}$ is identical to that
for the marginally bound critical 
particle $E^{2}=m^{2}$ obtained in Sec.~\ref{subsec:belt} and hence
we obtain
\begin{equation*}
\sin\theta\ge \sqrt{\frac{2}{3}}.
\end{equation*}
Thus, we can extend the original scenario
by Ba\~nados, Silk and West~\cite{BSW2009}
from the equator up to the latitude 
$\pm \mbox{acos} \sqrt{2/3}\simeq \pm 35.26^{\circ}$.

Moreover, we can also extend the analysis 
to include both marginally bound and unbound particles.
Also in this case, as seen in Eq.~(\ref{eq:R_critical_extremal}), 
the geodesic motion of the critical particle from infinity to 
the horizon is allowed if and only if 
$E^{2}\ge m^{2}$ and $(3E^{2}-m^{2})M^{2}\ge {\cal Q}$.
In other words, the condition obtained in Sec.~\ref{subsec:belt} 
also applies to the direct collision from infinity for both marginally bound
and unbound particles. So the upper limit on the latitude for an
arbitrarily high CM energy rises up to  
$\pm \mbox{acos}(\sqrt{3}-1)\simeq \pm 42.94^{\circ}$
as the energy of the particle is increased to infinity.
Therefore, Fig.~\ref{fg:high-velocity_collision_belt}
still applies if the original scenario by 
Ba\~nados, Silk and West~\cite{BSW2009}
is generalized to nonequatorial 
motion.

We have proven that the consideration of the global behavior does 
not change the condition for an arbitrarily high CM energy 
for the marginally bound and unbound critical particles in 
the Kerr black hole. However, it will not necessarily 
be true in more general black hole spacetimes.
\section{Conclusion and Discussion}
We have presented an expression for the CM energy of two general geodesic 
particles around a Kerr black hole. This is the generalization of the
formula obtained in the  
previous paper~\cite{Harada_Kimura2011} of the present authors, where 
the analysis was restricted to two massive geodesic particles 
of the same rest mass moving on the equatorial plane.
Applying this general expression,
we have shown that an unboundedly high 
CM energy can be realized only 
in the limit to the horizon and derived 
a formula for the CM energy for the near-horizon collision 
of two general geodesic 
particles.
Then, we have written down
the necessary and sufficient condition for an unboundedly high CM energy 
explicitly in terms of the conserved quantities of each particle  
and found that 
this reduces to that 
the ratio $(E_{1}-\Omega_{H}L_{1})/(E_{2}-\Omega_{H}L_{2})$ 
is infinitely large or infinitely close to zero
for the energy $E_{i}$ and angular momentum $L_{i}$ 
of particle $i$ ($i=1,2$).
Such a collision is possible at any latitude for any Kerr black hole with 
$0<a\leq M$ if the angular momentum is fine-tuned through 
multiple scattering in the vicinity of the horizon.

However, if we
concentrate on the direct collision scenario and the LSO collision scenario, 
the black hole in the limiting case must be maximally rotating 
to obtain an unboundedly high CM energy. 
Then, we find that the collision with an unboundedly high CM
energy can occur only on the belt between
latitudes $\pm 35.26^{\circ}$
if we only allow the  
bound and marginally bound critical massive particles
and $\pm 42.94^{\circ}$
if we allow all the possible critical particles.
This also applies to the original scenario proposed 
by Ba\~nados, Silk and West~\cite{BSW2009}.
It is suggested
that the collision with a very high CM energy might have
observational consequences in the contexts of the annihilation of dark matter 
particles~\cite{BSW2009,Banados_etal2010,Williams2011},
the high-energy hadron collision at inner edge of the accretion disks and 
the high-velocity collision of the compact objects around supermassive
black holes~\cite{Harada_Kimura2011}.
The present result strongly suggests that if signals
due to high-energy collision are to be observed, 
such signals can be produced primarily  
on the high-velocity collision belt centered at the equator 
of a (nearly) maximally rotating 
black hole but not from the polar regions. 

We briefly discuss the possible limitations of 
our result under the test particle approximation. Because of this 
approximation, we have neglected the self-gravity and back reaction
effects. In fact, these effects on particles 
orbiting a Kerr black hole have not been fully
understood yet. 
These effects are negligible and the inspiral will be 
always adiabatic if the mass ratio $\eta\equiv m/M$ 
is sufficiently small because these effects first
appear at $O(\eta)$. On the other hand, when $\eta $ is small but nonzero 
finite,
the back reaction effects due to a single high-velocity 
collision would considerably reduce the spin 
of the black hole~\cite{Berti_etal2009}.
It is also discussed that infinite collision energy is 
attained at the horizon after an infinite proper 
time and radiative 
effects cannot be neglected for such near-critical 
particles~\cite{Jacobson_Sotiriou2010,Berti_etal2009}. 
The effects of radiation reaction and conservative self-gravity
on the ISCO and LSO of a Kerr black hole are studied~\cite{Sundararajan2008,Ori_Thorne2000,Kesden2011,Barack_Sago2009,Barack_Sago2010,Barack_Sago2011}.
Based on those studies, these effects
are argued on the near-critical particles around 
a near-maximally rotating black hole in a different 
context~\cite{Barausse2010}. 
We speculate that these effects should be responsible for bounding the 
CM energy of the near-horizon collision. This is supported by the 
fully exact analysis of a system of charged spherical shells
surrounding an extremal Reissner-Nordst\"om black hole~\cite{Kimura_etal2011}.
It is clearly important to evaluate the upper bound 
in terms of the mass ratio $\eta$ for the 
collision of particles on the equatorial plane.
It will be the next step to study these effects on the collision
of general particles in the present context.

Here, we discuss the possible extension of the present analysis. 
Since $E-\Omega_{H} L =-\chi^{a}p_{a}$ for the horizon-generating Killing vector $\chi^{a}=\xi^{a}+\Omega_{H}\psi^{a}$ in the Kerr spacetime, 
we might extend the present analysis for the 
Kerr spacetime to more general stationary and axisymmetric 
spacetimes which admit a Killing vector $\chi^{a}$ and 
a Killing horizon ${\cal H}$, which is defined as a null 
hypersurface on which the Killing vector $\chi^{a}$ is also 
null. 
It is clear that the present analysis applies 
in a straightforward manner
if the analysis is restricted on the equatorial plane (e.g.~\cite{Zaslavskii2010_charged,WLGF2010,Zaslavskii2010_rotating}).
For the general geodesic orbits, the present analysis is still 
applicable only if the spacetime possesses three 
constants of motion and the geodesic equations can be written 
in the first-order form.
Note, however, that there is no analogue of the Carter constant 
for more general stationary and axisymmetric 
spacetimes (e.g.~\cite{Vigeland_etal2011}) and 
the present analysis will not immediately apply to 
general stationary and axisymmetric black holes.

Moreover, we may speculate that an arbitrarily high CM energy 
can be attained for the near-horizon collision even in the spacetime 
which is not stationary and axisymmetric but admits 
a Killing horizon ${\cal H}$ associated with a Killing vector $\chi^{a}$.
For a general geodesic particle, the quantity ${\cal A}=-\chi^{a}p_{a}$ 
is conserved. ${\cal A}$ must be positive in the vicinity of the 
horizon if $\chi^{a}$ is future-pointing timelike there.
This is the case for the nonmaximally rotating Kerr black holes. 
In such a case, it is 
clear that the critical particle, which has ${\cal A}=0$, cannot approach the horizon from outside. On the other hand, for the maximally rotating Kerr 
black hole, this may not apply and this is exactly what Ba\~nados, Silk and West~\cite{BSW2009} exploit. Now, we conjecture that if and only 
if particles 1 and 2 collide near the Killing horizon and 
the ratio ${\cal A}_{1}/{\cal A}_{2}$ is infinitely large or 
infinitely close to zero, 
the CM energy of the two particles is unboundedly high possibly under some genericity condition.

\acknowledgments
The authors thank R.~Takahashi, K.~Nakao, N.~Sago, T.~Shiromizu, U.~Miyamoto, M.~Saijo, J.~Novak, T.~Igata and E.~Barausse 
for very helpful comment. 
T.H. was supported by a Grant-in-Aid for Scientific Research from the Ministry of Education, Culture, Sports, Science and Technology of Japan 
[Young Scientists (B) No. 21740190].

\end{document}